\documentclass[12pt]{article}
\usepackage{epsfig}
\usepackage{amsmath}
\usepackage{amsfonts}

\begin{document}
\pagestyle{empty}
\begin{flushright}
UMN-TH-2620/07\\
\end{flushright}
\vspace*{5mm}

\begin{center}
{\Large\bf Warped Phenomenology 
in the Holographic Basis}
\vspace{1.0cm}

{\sc Brian Batell}\footnote{E-mail:  batell@physics.umn.edu}
{\small and}
{\sc Tony Gherghetta}\footnote{E-mail:  tgher@physics.umn.edu}
\\
\vspace{.5cm}
{\it\small {School of Physics and Astronomy\\
University of Minnesota\\
Minneapolis, MN 55455, USA}}\\
\end{center}

\vspace{1cm}
\begin{abstract}
The holographic basis is a novel tool which allows for a quantitative description of elementary/composite mixing in holographic duals of warped models. We apply this tool to bulk fermions in a slice of AdS$_5$ and determine the precise admixture of elementary source field and composite resonances forming the Standard Model fermions. In particular, for the phenomenologically important case of an IR localized right-handed top quark, we show that the massless eigenstate is approximately a $50/50$ elementary-composite admixture. We also translate, in a simple yet quantitative manner, several of the phenomenological successes enjoyed by warped models to the language of partial compositeness.   
\end{abstract}

\vfill
\begin{flushleft}
\end{flushleft}
\eject
\pagestyle{empty}
\setcounter{page}{1}
\setcounter{footnote}{0}
\pagestyle{plain}

\section{Introduction}
Warped extra dimensions provide a novel framework in which to explore physics beyond the Standard Model (SM), offering natural solutions to many outstanding puzzles in particle physics. Of course the most well-known achievement is the geometrical solution to the hierarchy problem \cite{rs}, realized by separating gravity from the Higgs boson in the fifth dimension. By extending the SM fields into the bulk, it is possible to naturally generate Yukawa hierarchies, suppress flavor-changing-neutral-currents (FCNC), achieve gauge coupling unification, and construct realistic models of electroweak symmetry breaking with Kaluza-Klein (KK) excitations within reach of the LHC \cite{review}.  

Although the notion of a curved higher-dimensional spacetime may seem speculative, the utility of this framework is undeniable in light of the AdS/CFT correspondence \cite{ads1,ads2,ads3}, according to which these five-dimensional (5D) theories are dual to purely four-dimensional (4D) strongly coupled gauge theories. Typically in such theories, perturbative approaches are futile due to the strong coupling, and hence important observables are difficult to calculate. Since the 5D theory is weakly coupled, the warped dimension can be viewed as a calculational tool for the 4D strongly coupled theory.

Theories in a slice of anti-de Sitter (AdS) space have an especially rich holographic interpretation \cite{pheno1,pheno2,pheno3}.  The existence of an infrared (IR) brane in 5D translates to a spontaneous breakdown of conformal symmetry in 4D. Accordingly, bound states built out of fundamental conformal field theory (CFT) fields appear. These composites are the ``holograms'' of the KK modes. An ultraviolet (UV) brane corresponds to adding an elementary degree of freedom, the ``source'' field, to the 4D theory which mixes with the composites. Hence, the true mass eigenstates of the theory contain some admixture of elementary (source) and composite (CFT) degrees of freedom. 

Recently we proposed a new tool, the {\it holographic basis} \cite{holo}, which can be used to quantify the mixing between the elementary and composite sectors. Rather than perform a Kaluza-Klein decomposition, we expand the bulk field directly in terms of purely source field and purely composite resonances. The effective theory contains contains in general both kinetic and mass mixing between the two sectors. It is straightforward to transform from the holographic basis to the KK basis, and this transformation tells us the precise elementary/composite content of the physical degrees of freedom in the 4D dual theory.

Ref. \cite{holo} analyzed only bosonic fields. In this paper, we formulate the holographic basis for bulk fermions in a slice of AdS$_5$. The holographic interpretation of warped fermions was given in Ref. \cite{fermholo}. One particularly striking aspect of the dual theory is that, in certain cases depending on the bulk fermion mass, an additional elementary degree of freedom marries with the source field via a mass mixing, while simultaneously the CFT produces an exponentially light mode. We will confirm this result explicitly with our formalism. As one of the main applications of the holographic basis, we compute the elementary/composite content of SM fermions in realistic warped models. In particular, we show the IR localized right-handed top quark is approximately a $50/50$ elementary-composite admixture.

Many aspects of the so-called holographic procedure (associating a CFT operator with a bulk field, constructing a boundary effective action, etc.) can seem mysterious, especially when compared with the more standard Kaluza-Klein method of analyzing 5D theories. In this regard, using the holographic basis is more in line with the standard approach in that we decompose the field in terms of a complete set of states and compactify the theory. Since these tools are familiar, we hope the holographic basis will allow for a more accessible (and quantitative) approach to understanding holography. 

To this end, we use the holographic basis to recast several important aspects of warped phenomenology into the language of strongly coupled composites mixing with an elementary sector. Specifically, we consider the holographic interpretations of the Yukawa hierarchies, Randall-Sundrum GIM mechanism, and the custodial protection in electroweak precision tests \cite{custodial}, which become quite transparent with the help of the holographic basis. A phenomenological approach to partial compositeness has also been considered in Ref.~\cite{crss}.

We begin in Section 2 by reviewing fermions in warped space and their holographic interpretation. The holographic basis for fermions is proposed in Section 3, where we specify the bulk profiles of the elementary and composite states, set up the general eigenvalue problem, and outline how to transform to the mass  eigenbasis. In Section 4, we use the holographic basis to compute the elementary/composite content of SM fermions in warped models. Section 5 is devoted to understanding the holographic picture of several robust features of bulk RS models. Our conclusions are presented in Section 6.

\section{Warped/composite fermions}

We set the stage for the holographic basis by recalling aspects of bulk fermions in a slice of AdS$_5$ \cite{fermions,gp} and their holographic interpretation \cite{fermholo}. The metric describing 5D anti-de Sitter space is 
\begin{equation}
ds^2=e^{-2ky}\eta_{\mu\nu}dx^\mu dx^\nu+dy^2,
\end{equation}
where $k$ is the AdS$_5$ curvature scale. The interval ranges from $y=0$ to $y=\pi R$ where there exists a UV and IR brane, respectively. We denote 5D indices with Latin letters ($A, B, \dots$) and 4D indices with Greek letters ($\mu,\nu,\dots$). We use the flat metric $\eta ={\rm diag}(-,+,+,+)$ to raise and lower 4D indices.

The action for a Dirac fermion $\Psi(x,y)$ propagating in the bulk is
\begin{equation}
S_{bulk}=\int d^5 x\sqrt{-g} \left[\frac{1}{2}\overline\Psi e^M_A \Gamma^A D_M \Psi-\frac{1}{2}D_M\overline\Psi e^M_A \Gamma^A \Psi+ c k \overline\Psi \Psi\right],
\label{a1}
\end{equation}
where $e^M_A$ is the vielbein, $D_M=\partial_M +\omega_M$ is the covariant derivative. The spin connection pieces $\omega_M$ cancel in (\ref{a1}). 
Defining $\psi=e^{-2 k y}\Psi$, and the two-component spinors $\gamma^5\psi_\pm= \pm \psi_\pm$,
we can write the action as
\begin{eqnarray}
S_{bulk}&=&\int d^5x\left[e^{ky} \overline\psi_+\gamma^\mu\partial_\mu\psi_+ -\frac{1}{2}\overline\psi_+ (\partial_5-ck)\psi_- +\frac{1}{2}(\partial_5+ck)\overline\psi_+ \psi_- \right.\nonumber\\
& &\left.+ e^{ky}\overline\psi_-\gamma^\mu\partial_\mu\psi_- +\frac{1}{2}\overline\psi_- (\partial_5+c k)\psi_+ -\frac{1}{2}(\partial_5-c k)\overline\psi_- \psi_+ \right].
\label{a2}
\end{eqnarray}

Special attention must be paid to boundary terms that arise when varying the action. Usually when dealing with 5D theories, one of the fields $\psi_+$ or $\psi_-$ is taken to have a Dirichlet condition at the boundaries, guaranteeing that all boundary terms vanish and allowing integration by parts freely. But if more general boundary conditions are applied to $\psi_\pm$\footnote{
For an in-depth analysis of general fermion boundary conditions see Ref. \cite{fermionbc}.}, as occurs when using the holographic procedure \cite{fermholo}, the variational principle is not satisfied due to nonvanishing boundary terms.  As we will see in Sec. 3. the profiles defining the holographic basis satisfy nontrivial boundary conditions, and therefore generic boundary terms will not vanish. To satisfy the variational principle with the holographic basis, we will require an additional boundary term:
\begin{equation}
S_{boundary}=\frac{1}{2}\int d^4 x\Big[~\overline\psi_+ \psi_- + \overline\psi_- \psi_+ \Big]\bigg\vert_0^{\pi R},
\label{boundary}
\end{equation}
which upon variation, cancels the boundary terms arising from the bulk action. The total action $S=S_{bulk}+S_{boundary}$ satisfies the variational principle. Note that the additional term (\ref{boundary}) automatically vanishes if one of the fields satisfies the typical Dirichlet condition.

Finally we integrate the second and sixth terms in (\ref{a2}) by parts to write the total action as
\begin{equation}
S=\int d^5x \Big[e^{ky} \overline\psi_+\gamma^\mu\partial_\mu\psi_+ + e^{ky}\overline\psi_-\gamma^\mu\partial_\mu\psi_- +(\partial_5+c k)\overline\psi_+ \psi_- +\overline\psi_- (\partial_5+c k)\psi_+\Big].
\label{a3}
\end{equation}
Again, when we work in the holographic basis, the boundary terms that arise in the integration by parts are canceled by (\ref{boundary}). This will be our starting point for analyzing the elementary/composite mixing in the holographic basis.

\subsection{Kaluza-Klein basis}
The traditional method of analyzing 5D theories is to perform a Kaluza-Klein decomposition. We expand the 5D fields $\psi_\pm(x,y)$ in a complete set of states
\begin{equation}
\psi_\pm(x,y)=\sum_{n=0}^\infty \psi^n_\pm(x)f^n_\pm(y),
\label{kkexp}
\end{equation}
where the eigenfunctions satisfy the equations of motion
\begin{equation}
\big[\partial_5\pm c k\big]f^n_\pm(y)=\pm m_n e^{k y}f^n_\mp(y),
\label{eom1}
\end{equation}
and the orthonormal condition
\begin{equation}
\int_0^{\pi R}dy ~ e^{ky} f^n_\pm(y) f^m_\pm(y)=\delta^{nm}.
\label{ortho}
\end{equation}
We will be interested in the case where a massless chiral mode exists in the theory. Without loss of generality, we will take $\psi_+(x,y)$ to have a zero mode. The boundary conditions allowing the existence of a zero mode $\psi^0_+(x)$, with $\psi^0_-(x)$ projected out, are
\begin{eqnarray}
(\partial_5+ck)f^n_+(y)\bigg\vert_{0,\pi R}&=&0, \label{kkbc1} \\
f^n_-(y)\bigg\vert_{0,\pi R}&=&0.
\label{kkbc2}
\end{eqnarray}
The normalized zero mode wavefunction is given by 
\begin{equation}
f^0_+(y)=\sqrt{\frac{(1-2c)k}{e^{(1-2c)\pi k R}-1}} e^{-c k y}.
\label{0mode}
\end{equation}
The mass eigenvalues are determined by applying boundary conditions (\ref{kkbc1}) and (\ref{kkbc2}) to the solutions of (\ref{eom1}), which yields the equation determining the spectrum: 
\begin{equation}
J_{c - \frac{1}{2}}\left(\frac{m_n}{k}\right)Y_{c- \frac{1}{2}}\left(\frac{m_n}{k}e^{\pi k R}\right)
-Y_{c- \frac{1}{2}}\left(\frac{m_n}{k}\right)J_{c- \frac{1}{2}}\left(\frac{m_n}{k}e^{\pi k R}\right)=0.
\label{spectrum1}
\end{equation}
Explicitly, the spectrum contains a chiral zero mode $\psi_+^0(x)$ and a tower of Dirac fermions $\psi^n(x)^{\rm T}=(\psi_+^n(x),\psi_-^n(x))$ with mass $m_n$. The Kaluza-Klein basis is, by construction, diagonal in field space, with no quadratic mixing in the effective Lagrangian, and is therefore the physical mass eigenbasis. In contrast, we will see in the holographic basis that there is quadratic mixing between an elementary sector and a set of composite resonances. The physical states therefore contain a mixture of elementary and composite states. We elaborate on this next with a discussion of the AdS/CFT inspired holographic interpretation of bulk fermions.  
 
\subsection{The holographic theory}

The theory of bulk fermions in a slice of AdS$_5$ can be given a holographic interpretation \cite{fermholo}, in which the physical states of the theory contain a mixture of elementary and composite states. Ultimately our goal is to quantify this mixing, which we will do in the next section. Here we focus on several main results of \cite{fermholo} which will provide motivation for the holographic basis. 

To implement the holographic procedure, we must first specify which bulk field $\psi_\pm(x,y)$ will correspond to the elementary ``source'' field in the dual theory. We are interested in the holographic dual of the theory with a chiral zero mode described in Section 2.1, so clearly we should choose $\psi_+$ as our source field, with arbitrary UV boundary value $\widetilde{\psi}_+(x)$. Then by the AdS/CFT correspondence, there is a corresponding CFT operator ${\cal O}_-$ which is sourced by $\widetilde{\psi}_+(x)$ in the 4D generating functional of the dual theory. We can compute the two-point correlator which is contained in the self-energy $\Sigma(p)$ \cite{fermholo}:
\begin{equation}
\Sigma(p)=\frac{1}{g_5^2}~\frac{p}{\!\not\!p}~
\frac{J_{c - \frac{1}{2}}\left(\frac{i p}{k}\right)Y_{c- \frac{1}{2}}\left(\frac{i p}{k}e^{\pi k R}\right)
-Y_{c- \frac{1}{2}}\left(\frac{i p}{k}\right)J_{c- \frac{1}{2}}\left(\frac{i p}{k}e^{\pi k R}\right)}
{J_{c + \frac{1}{2}}\left(\frac{i p}{k}\right)Y_{c- \frac{1}{2}}\left(\frac{i p}{k}e^{\pi k R}\right)
-Y_{c + \frac{1}{2}}\left(\frac{i p}{k}\right)J_{c- \frac{1}{2}}\left(\frac{i p}{k}e^{\pi k R}\right)}.
\end{equation}
The poles in the correlator, which are the zeros of the function in the denominator, correspond to the diagonal masses $M_n$ (distinct from the physical masses $m_n$) of the composite CFT states:
\begin{equation}
J_{c + \frac{1}{2}}\left(\frac{M_n}{k}\right)Y_{c- \frac{1}{2}}\left(\frac{M_n}{k}e^{\pi k R}\right)
-Y_{c + \frac{1}{2}}\left(\frac{M_n}{k}\right)J_{c- \frac{1}{2}}\left(\frac{M_n}{k}e^{\pi k R}\right)=0.
\label{spectrum2}
\end{equation}
Notice that this mass spectrum would arise if we changed the boundary conditions in (\ref{kkbc1}) and
(\ref{kkbc2}) from Neumann to Dirichlet at the UV boundary for $\psi_+(x,y)$, and vice-versa for $\psi_-(x,y)$. This observation will be crucial in defining the holographic basis. 

The self-energy $\Sigma(p)$ also gives us information about the nature of the elementary source field. For $c>-1/2$, the source is chiral since no massless pole appears in either the high or low momentum expansion of $\Sigma(p)$. However, for $c<-1/2$, the high-energy expansion yields
\begin{equation}
\Sigma(p)\simeq - \frac{1}{g_5^2 k } i \!\not\!p ~\left[ \frac{(1+2c)k^2}{p^2}+\frac{1}{3+2c}+\dots\right].
\label{sigpole}
\end{equation}
The second term above corresponds to a kinetic term for the source. The first term is a massless pole. This pole cannot be attributed to a composite particle built out of CFT fields, since in the high momentum regime the conformal symmetry is unbroken. The correct interpretation is that a new elementary degree of freedom appears in the theory and marries with the source field via a mass mixing on this branch. The source mass can be extracted from the correlator and is
\begin{equation}
\widetilde{M}=k\sqrt{(3+2c)(1+2c)}.
\label{M0}
\end{equation}
This mass is defined near the UV scale and, due to radiative corrections involving CFT interactions, will be modified as we run down to the IR scale.

Notice that as the source field becomes massive, the CFT produces an exponentially light mode. Expanding (\ref{spectrum2}) for $p\ll k e^{-\pi k R}$ for values $c< - 1/2$, we find that
\begin{equation}
M_1\simeq \sqrt{4c^2-1}k\,e^{-(\frac{1}{2}-c)\pi k R}.
\label{lightcft}
\end{equation}
We might then expect that the massless eigenstate is primarily composed of this composite state, which indeed is the case for $c<-1/2$. On the other hand, for $c>-1/2$, the source field is chiral and no light composite mode exists, indicating that the massless eigenstate is primarily elementary.

The scaling dimension of the operator ${\cal O}_-$ can also be extracted from $\Sigma(p)$, and is given by
\begin{equation}
\Delta_{-} = \frac{3}{2}+\bigg\vert c+\frac{1}{2} \bigg\vert.
\label{dimo}
\end{equation}
This tells us when the source/CFT mixing is relevant, marginal, or irrelevant in terms of the bulk mass 
parameter $c$.

Because of the source-CFT interaction $(\widetilde{\overline{\psi}}_+ {\cal O}_- +{\rm h.c.})$, the physical spectrum is given by (\ref{spectrum1}) rather than (\ref{spectrum2}). This can be viewed as a result of quadratic mixing between the elementary source field and the composite resonances in the effective Lagrangian. We now move on to our primary goal of a quantitative description of this elementary/composite mixing.

\section{The holographic basis}

We are now ready to propose the holographic basis for fermions. The idea is very simple: we expand the bulk fields $\psi_\pm(x,y)$ in a set of 4D fields corresponding to a purely elementary sector and a tower of CFT resonances.  Consider the following alternate expansion of the bulk fields:
\begin{eqnarray}
\psi_+(x,y)&=&\psi^s(x)g^s(y)+\sum^\infty_{n=1}\lambda^n_+(x)g^n_+(y),\label{holo+}\\
\psi_-(x,y)&=&\chi(x)g^\chi(y)+\sum^\infty_{n=1}\lambda^n_-(x)g^n_-(y), \label{holo-}
\end{eqnarray}
where $\psi^s(x)$ corresponds to the elementary source field, the $\lambda^n_\pm(x)$ correspond to the CFT bound states, and $\chi(x)$ is an additional elementary degree of freedom external to the CFT.

Obviously the holographic profiles $g^s(y), g^\chi(y), g^n_\pm(y)$ must differ from the KK profiles $f^n_\pm$ if we are to describe holographic mixing. Let us first discuss the composite profiles $g^n_\pm(y)$. As we pointed out in the previous section, the CFT resonances have diagonal masses given by the spectrum (\ref{spectrum2}). Hence, the profiles $g^n_\pm(y)$ must satisfy the bulk equation (\ref{eom1}) with $m_n$ replaced by $M_n$, as well as the following holographic boundary conditions:
\begin{eqnarray}
g^n_+(y)\bigg\vert_0=0, & (\partial_5+ck)g^n_+(y)\bigg\vert_{\pi R}=0, \nonumber \\
(\partial_5-ck)g^n_-(y)\bigg\vert_{0}=0, & g^n_-(y)\bigg\vert_{\pi R}=0.
\label{holobc} 
\end{eqnarray}
The CFT wavefunctions are therefore given by
\begin{equation}
g^n_\pm(y)=N^{CFT}_n e^{\frac{ky}{2}}\Big[J_{c\pm \frac{1}{2}}\left(\frac{M_n}{k}e^{ky}\right)+\kappa(M_n)Y_{c\pm \frac{1}{2}}\left(\frac{M_n}{k}e^{ky}\right)\Big],
\label{gCFT}
\end{equation}
with the coefficient $\kappa(M_n)$ determined from the holographic boundary conditions (\ref{holobc}):
\begin{equation}
\kappa(M_n)=-\frac{J_{c+ \frac{1}{2}}\left(\frac{M_n}{k}\right)}{Y_{c+ \frac{1}{2}}\left(\frac{M_n}{k}\right)}=-\frac{J_{c- \frac{1}{2}}\left(\frac{M_n}{k}e^{\pi k R}\right)}{Y_{c- \frac{1}{2}}\left(\frac{M_n}{k}e^{\pi k R}\right)}.
\label{kappa}
\end{equation}
We see explicitly that (\ref{kappa}) is equivalent to (\ref{spectrum2}).
We also give the normalization $N_n^{CFT}$ (chosen so the resonance kinetic term is canonical) which is useful when computing the mixing coefficients in the holographic Lagrangian:
\begin{equation}
N_n^{CFT}=\frac{\pi M_n}{\sqrt{2k}}\frac{Y_{c+ \frac{1}{2}}\left(\frac{M_n}{k}\right)Y_{c- \frac{1}{2}}\left(\frac{M_n}{k}e^{\pi k R}\right)}
{\sqrt{Y^2_{c + \frac{1}{2}}\left(\frac{M_n}{k}\right)-Y^2_{c - \frac{1}{2}}\left(\frac{M_n}{k}e^{\pi k R}\right)}}.
\end{equation}

Now consider the source wavefunction. In analogy with the scalar case \cite{kw}, the AdS/CFT correspondence prescribes the bulk field $\psi_+(x,y)$ to display the following asymptotic behavior in the UV in order to construct the proper boundary action:
\begin{equation}
\psi_+(x,y)\rightarrow e^{(2-\Delta_-)ky}\widetilde{\psi}_+(x)+\cdots,
\end{equation}
where $\Delta_-$ is the scaling dimension of the operator ${\cal O}_-$ given in (\ref{dimo}). The field $\widetilde{\psi}_+(x)$ is the source field and is equivalent to $\psi^s(x)$ appearing in (\ref{holo+}) up to an overall normalization. We therefore postulate as our source wavefunction 
\begin{equation}
g^s(y)=N_s e^{(2-\Delta_-) k y}
=\begin{cases}
\sqrt{\frac{(1-2c)k}{e^{(1-2c)\pi k R}-1}}e^{-cky}  \quad~~~ {\rm for} \quad c>-\frac{1}{2}~,\\\\
 \sqrt{\frac{(3+2c)k}{e^{(3+2c)\pi k R}-1}}e^{(1+c)ky} \quad  {\rm for} \quad  c<-\frac{1}{2}~. 
\end{cases}
\label{gsource}
\end{equation}
The normalization $N_s$ renders the 4D action canonical. The source profile has a very simple holographic explanation. Consider the profile with respect to a flat metric, $\tilde{g}^s=e^{ky/2}g^s$, for large values of $|c|$, corresponding to irrelevant source-CFT mixing. In this case the source field is localized on the UV brane and thus has a very weak overlap with the composite CFT resonances, which are IR localized. On the other hand, for $-3/2<c<1/2$ where the source-CFT mixing is relevant, the source field is in fact localized on the IR brane and overlaps very strongly with the composites. This picture is identical to the one obtained by analyzing the operator dimension $\Delta_-$ (\ref{dimo}) \cite{fermholo}.  

The most difficult wavefunction to ascertain is $g^\chi(y)$ associated with the elementary field $\chi(x)$. The primary motivation for the form of this profile will be that it reproduces the correct mass mixing between $\psi^s(x)$ and $\chi(x)$ (\ref{M0}), as we will discuss in the next section. The correct wavefunction turns out to be
\begin{equation}
g^\chi(y)
=\begin{cases}
0 ~\quad\quad\quad\quad\quad\quad\quad~{\rm for} \quad c>-\frac{1}{2}~,\\\\
\sqrt{\frac{(1+2c)k}{e^{(1+2c)\pi k R}-1}}e^{cky} \quad  {\rm for} \quad  c<-\frac{1}{2}~. 
\end{cases}
\label{gchi}
\end{equation}
For $c>-1/2$, the elementary field $\chi(x)$ is absent from the theory, and furthermore, the source profile (\ref{gsource}) is identical to the zero mode profile (\ref{0mode}), which is clearly consistent with the source field being chiral. However, for $c<-1/2$, the external $\chi(x)$ marries with the source $\psi^s(x)$ via a mass mixing. As stated previously, this is seen in the correlator (\ref{sigpole}) by the presence of a new pole which cannot be attributed to the CFT as it is present at high energies when the conformal symmetry is unbroken.

\subsection{The holographic eigenvalue problem}
We now have the tools to quantitatively describe elementary/composite mixing in theories with bulk fermions. Inserting the holographic expansion (\ref{holo+}) and (\ref{holo-}) into the action (\ref{a3}), and integrating over $y$, we can derive the low energy action for the holographic theory:
\begin{equation}
S=S(\psi^s,\chi)+S(\lambda^n_\pm)+S_{mix},
\label{aholo}
\end{equation}
where 
\begin{eqnarray}
S(\psi^s,\chi)&=&\int d^4x \Big[\overline{\psi}^s\gamma^\mu\partial_\mu\psi^s+\overline{\chi}\gamma^\mu\partial_\mu \chi +M_s (\overline{\psi}^s\chi +\overline{\chi}\psi^s)\Big],\\
S(\lambda^n_\pm)&=& \sum_{n=1}^\infty\int d^4x \Big[  \overline{\lambda}^n_+ \gamma^\mu\partial_\mu\lambda^n_+ + \overline{\lambda}^n_- \gamma^\mu\partial_\mu\lambda^n_- +M_n (\overline{\lambda}^n_+\lambda^n_- +\overline{\lambda}^n_-\lambda^n_+)\Big], \\
S_{mix}&=& \sum_{n=1}^\infty \int d^4 x \Big[ z^s_n(\overline{\psi}^s \gamma^\mu\partial_\mu \lambda^n_+ + \overline{\lambda}^n_+\gamma^\mu\partial_\mu \psi^s)
+z^\chi_n(\overline{\chi} \gamma^\mu\partial_\mu \lambda^n_- + \overline{\lambda}^n_-\gamma^\mu\partial_\mu \chi)\nonumber \\
& &+\mu_n^s(\overline{\psi}^s \lambda^n_- +\overline{\lambda}^n_-\psi^s)+\mu_n^\chi(\overline{\chi}\lambda^n_+ +\overline{\lambda}^n_+\chi) \Big].
\end{eqnarray} 
The mixing coefficients $M_s, z_n^s, z_n^\chi, \mu_n^s,$ and $\mu_n^\chi$ are calculated from wavefunction overlap integrals:
\begin{eqnarray}
M_s&=&\int_0^{\pi R} dy~g^\chi(y) (\partial_5+ck)g^s(y),\label{msource}\\
z^s_n&=&\int_0^{\pi R} dy ~e^{ky}g^s(y)g^n_+(y),\label{kmixsource}\\
z^\chi_n&=&\int_0^{\pi R}dy~ e^{ky}g^\chi(y)g^n_-(y),\label{kmixchi}\\
\mu^s_n&=&\int_0^{\pi R}dy~g^n_-(y)(\partial_5+ck)g^s(y),\label{mmixsource}\\
\mu^\chi_n&=& M_n z^\chi_n. \label{mmixchi}
\end{eqnarray}
The nonvanishing kinetic mixing indicates that the holographic basis is not orthogonal.

The system (\ref{aholo}) can be compactly represented using matrices. Defining the field vector $\vec{\psi}^{\rm T}= (\psi^s,\chi,\lambda^1_+,\lambda^1_-, \lambda^2_+,\lambda^2_-,\dots)$, the Lagrangian is
\begin{equation}
{\cal L}=\vec{\overline\psi}^{\rm T} {\bf Z} \gamma^\mu \partial_\mu \vec{\psi}+\vec{\overline\psi}^{\rm T} {\bf M}\vec{\psi}.
\end{equation}
The mixing matrices are 
\begin{eqnarray}
{\bf Z}&=&  \left( \begin{array}{ccccccc} 
1      & 0       & z^s_1 & 0        & z^s_2 & 0        &\cdots \\ 
0      & 1       & 0     & z_1^\chi & 0     & z_2^\chi &\cdots \\
z^s_1  & 0       & 1     & 0        & 0     & 0        &\cdots \\
0      & z_1^\chi & 0     & 1        & 0     & 0        &\cdots \\
z^s_2  & 0       & 0     & 0        & 1     & 0        &\cdots \\
0      & z_2^\chi & 0     & 0        & 0     & 1        &\cdots \\
\vdots & \vdots  & \vdots& \vdots   &\vdots &\vdots    &\ddots 
\end{array} \right),\label{kmix}\\
{\bf M}&=&  \left( \begin{array}{ccccccc} 
0       & M_s        & 0          & \mu^s_1 & 0         & \mu^s_2  &\cdots\\ 
M_s     & 0          & \mu^\chi_1 & 0      & \mu^\chi_2 &  0       &\cdots\\
0       & \mu^\chi_1 & 0          & M_1    & 0          &  0       &\cdots\\
\mu^s_1 & 0          & M_1        & 0      & 0          & 0        &\cdots\\
0       & \mu^\chi_2 & 0          & 0      & 0          & M_2        &\cdots\\
\mu^s_2 & 0          &0           & 0      & M_2          & 0        &\cdots \\
\vdots  & \vdots     & \vdots     & \vdots & \vdots      &\vdots    &\ddots \end{array} \right).\label{mmix}
\end{eqnarray}
Note that for $c>-1/2$, the field $\chi(x)$ is absent from the system and the corresponding entries in the above matrices are to be removed.

Diagonalizing this system can be seen as a four-step field redefinition:
\begin{equation}
\vec{\psi}\rightarrow {\bf R} {\bf V}{\bf T}^{-1}{\bf U}\vec{\psi}.
\label{transform}
\end{equation}
Let us discuss the role of each of the matrices in turn. First, we diagonalize the kinetic matrix ${\bf Z}$ with a field rotation ${\bf U}$. The kinetic energies are then diagonal, but not canonical, and thus need to be rescaled by the matrix ${\bf T}={\rm diag}(1/\sqrt{{\rm eigenvalue}({\bf Z})})$.
Next, we diagonalize the (rotated and scaled) mass mixing matrix,
\begin{equation}
{\bf M'}={\bf T}{\bf U}{\bf M}{\bf U}^{\rm T}{\bf T},
\end{equation}
with an additional field rotation ${\bf V}$. This leaves the mass matrix in the form ${\bf V}{\bf M'}{\bf V}^{\rm T}={\rm diag}(0, 0, m_1, -m_1, m_2,-m_2,\dots)$. The action is put in the correct form, with block 
off-diagonal masses, by the matrix ${\bf R}$:
\begin{eqnarray}
{\bf R}&=& \frac{1}{\sqrt{2}} \left( \begin{array}{rrrrr} 
1  & 1    & 0     & 0      &\cdots \\ 
1   & -1    & 0     & 0      &\cdots \\
0   & 0    & 1    & 1       &\cdots \\
0   & 0    & 1     & -1       &\cdots \\
\vdots & \vdots  & \vdots& \vdots     &\ddots 
\end{array} \right).\label{R}
\end{eqnarray}
If the field $\chi(x)$ is absent from the system, this matrix is modified: the entries corresponding to $\chi$ are removed and the source entry is changed from $1 / \sqrt{2} \rightarrow 1$.  

Although we have not been able to diagonalize this system analytically, it is straightforward to analyze a truncated system numerically for a given set of parameters $(c, k, R)$. The diagonalized mass eigenvalues from the truncated system match the true spectrum (\ref{spectrum1}) more and more precisely as we increase the number of composite states in the system, telling us that indeed the holographic basis is correctly describing the dual theory. For more details and numerical evidence for the bosonic case, see Ref. \cite{holo}. We now give analytic expressions for the mixing coefficients (\ref{msource})-(\ref{mmixchi}). 

\subsubsection{$c>-\frac{1}{2}$}

In the case $c>-1/2$, relevant for nearly all phenomenological purposes, there exists only kinetic mixing  since $\chi(x)$ is absent from the theory and since $(\partial_5+ck)g^s(y)=0$. We only have to deal with the kinetic mixing $z_n^s$, shown in (\ref{kmixsource}). Inserting the source (\ref{gsource}) and CFT (\ref{gCFT}) wavefunctions, we have
\begin{eqnarray}
z_n^s&=&N_sN_n^{CFT}\int_0^{\pi R} dy ~e^{\left(\frac{3}{2}-c\right)ky} \left[J_{c+ \frac{1}{2}}\left(\frac{M_n}{k}e^{ky}\right)+\kappa Y_{c+\frac{1}{2}}\left(\frac{M_n}{k}e^{ky}\right)\right], \nonumber \\ 
& =& \frac{N_s N_n^{CFT}}{k}\left(\frac{k}{M_n}\right)^{\frac{3}{2}-c}
\int_{u_0}^{u_{\pi R}} du ~ u^{\frac{1}{2}-c} \Big[J_{c+ \frac{1}{2}}(u)+\kappa Y_{c+\frac{1}{2}}(u)\Big], \nonumber \\
&=&-\frac{2 k N_s N_n^{CFT}}{\pi M_n^2 Y_{c+\frac{1}{2}}\left(\frac{M_n}{k}\right)},
\label{zns1}
\end{eqnarray}
where in the second line we have changed variables to $u=M_n e^{k y}/k$ and used Eq. (\ref{kappa}) to simplify the final result. 

\subsubsection{$c<-\frac{1}{2}$}
In this case, the existence of the additional elementary field $\chi(x)$ creates a more complicated mixing. The first effect of $\chi(x)$ is to marry with the source $\psi^s(x)$ through a mass mixing (\ref{msource}): 
\begin{equation}
M_s
= k\sqrt{\frac{e^{(1+2c)\pi k R}-1}{e^{(3+2c)\pi k R}-1}}\sqrt{(3+2c)(1+2c)},
\label{sourcemass}
\end{equation}
where we have used (\ref{gsource}) and (\ref{gchi}). Comparing this with (\ref{M0}), we see that an additional exponential factor appears in $M_s$. Of course, this is the low-energy value of the elementary mass mixing, and the extra coefficient should result from renormalization group running involving CFT insertions. The fact that (\ref{sourcemass}) is the correct low-energy mass mixing can be seen numerically by diagonalizing the system (\ref{aholo}) and reproducing the physical spectrum (\ref{spectrum1}). We have checked this explicitly for many cases.

There is also nontrivial kinetic mixing, (\ref{kmixsource}) and (\ref{kmixchi}), which can be calculated similarly to (\ref{zns1}):
\begin{eqnarray}
z_n^s&=&\frac{2 k N_s N_n^{CFT}}{\pi M_n^2}\left[\frac{(1+2c)k e^{(c-\frac{1}{2})\pi k R}}{ M_n Y_{c-\frac{1}{2}}\left(\frac{M_n}{k}e^{\pi k R}\right)}-\frac{1}{Y_{c+\frac{1}{2}}\left(\frac{M_n}{k}\right)}\right],\label{zns2} \\
z_n^\chi&=&\frac{2 k N_\chi N_n^{CFT}e^{(c-\frac{1}{2})\pi k R}}{\pi M_n^2 Y_{c-\frac{1}{2}}\left(\frac{M_n}{k}e^{\pi k R}\right)}.\label{znchi2}
\end{eqnarray}
Furthermore, there is elementary/composite mass mixing. After computing (\ref{znchi2}), the mass mixing $\mu_n^{\chi}$ is simply given by  (\ref{mmixchi}). There is also a mass mixing $\mu_n^s$ (\ref{mmixsource}) which can be computed using the wavefunctions (\ref{gCFT}) and (\ref{gsource}):
\begin{equation}
\mu^s_n
=(1+2c)k\frac{N_s}{N_\chi}z_n^{\chi},
\end{equation}
and thus can also be written in terms of $z_n^\chi$.

\subsection{Eigenvectors}
It is possible to obtain the eigenvectors directly by using orthogonality properties of the eigenfunctions. We simply equate the Kaluza-Klein (\ref{kkexp}) and holographic expansions (\ref{holo+}),(\ref{holo-}): 
\begin{eqnarray}
\sum_{n=0}^\infty \psi^n_+(x)f^n_+(y)&=&\psi^s(x)g^s(y)+\sum^\infty_{n=1}\lambda^n_+(x)g^n_+(y),\label{holo2+}\\
\sum_{n=1}^\infty \psi^n_-(x)f^n_-(y)&=&\chi(x)g^\chi(y)+\sum^\infty_{n=1}\lambda^n_-(x)g^n_-(y). \label{holo2-}
\end{eqnarray}

Using the orthonormal condition (\ref{ortho}), we can compute the precise linear combination of elementary and composite fields forming the mass eigenstates. For instance,
\begin{equation}
\psi^n_+(x)=v^{ns}_+\psi^s(x)+\sum^\infty_{m=1}v^{nm}_+\lambda^m_+(x),
\end{equation}
where
\begin{eqnarray}
v^{ns}_+ & = &\int_0^{\pi R} dy\, e^{ky} f^n_+(y) g^s(y)~, \label{v1}\\
v^{nm}_+ & = &\int_0^{\pi R} dy\, e^{ky} f^n_+(y) g^m_+(y)~. \label{v2}
\end{eqnarray}
Similar expressions can of course be written for $\psi_-(x,y)$.

Let us examine these expressions for $\psi_+$ in a bit more detail. Consider first the case $c>-1/2$. In this case, the zero mode profile $f^0(x)$ is identical to the source profile $g^s(x)$ (see (\ref{0mode}) 
and (\ref{gsource})), and the eigenvector for the massless chiral mode takes the simple form:
\begin{equation}
\psi^0_+(x)=\psi^s(x)+\sum^\infty_{n=1}z_n^s\lambda^n_+(x),
\label{ev-}
\end{equation}
where $z_n^s$ is computed in (\ref{zns1}). The massless mode is primarily elementary in this case, since $z_n^s<1$. The massive modes (KK states) are purely composite since, again, the source and zero mode profiles are equivalent, so by orthogonality (\ref{ortho}) $v_+^{ns}=0$.

Now, on the other branch $c<-1/2$, the mixing changes drastically. The zero mode and source profiles are different for these values of $c$. However, the eigenvectors can still be computed analytically. Consider first the elementary content of $\psi^0_+(x)$:
\begin{eqnarray}
v^{0s}_+&=&\frac{1}{2}\sqrt{\frac{(1-2c)}{e^{(1-2c)\pi k R}-1}}\sqrt{\frac{(3+2c)}{e^{(3+2c)\pi k R}-1}}(e^{2 \pi k R}-1)~,\\
&\simeq&\begin{cases}
\sqrt{(\frac{1}{2}-c)(\frac{3}{2}+c)} \quad\quad\quad\quad\quad~{\rm for} \quad -\frac{3}{2}<c<-\frac{1}{2}~,\\\\
\sqrt{(c-\frac{1}{2})(c+\frac{3}{2})}e^{-|3/2+c|\pi k R}   \quad  {\rm for} \quad  c<-\frac{3}{2}~.
\end{cases}
\label{v0+}
\end{eqnarray}
This is precisely what we would guess from analyzing the operator dimensions. For $-3/2<c<-1/2$ there is relevant mixing between the elementary and composite sectors, and so the zero mode contains a significant elementary component. For $c<-3/2$, however, the source-CFT mixing is irrelevant and the zero mode contains only an exponentially small elementary component. 

In fact, the massless mode in this case is almost purely the first composite CFT state, as we guessed previously from the CFT spectrum, which contains a very light mode. For $c<-3/2$ we find for the first composite state $v^{01}_+ \sim  - 1$, while all other $v^{0n}_+$ are exponentially suppressed. 

The situation also changes for the massive modes $\psi^n_+$ for $c<-1/2$, which now become partly elementary, $v^{ns}_+\neq0$. Although we won't analyze the $\psi^n_-$ eigenvectors in any detail, it is worth mentioning that the massive KK modes behave in a similar way. For $c>-1/2$ the massive eigenstates are purely composite ($\chi(x)$ is absent from the theory), while for $c<-1/2$ the states contain the elementary field $\chi(x)$.

By a similar use of the holographic wavefunctions, one can write the source and composite fields in terms of the mass eigenstates (inverse transformation). 
For the fields $\psi^s(x)$ and $\lambda^n_+(x)$, we obtain
\begin{eqnarray}
\psi^s(x)&=&\sum_{n=0}^\infty \omega^{sn}_+ \psi^n_+(x), \label{sinv} \\
 \lambda^n_+(x)&=&\sum_{m=0}^\infty \omega^{nm}_+   \psi^m_+(x) \label{laminv} ,
\end{eqnarray}
where the coefficients are given by
\begin{eqnarray}
\omega^{sn}_+ &=& \frac{f^n_+(0)}{g^s(0)}, \\
\omega^{nm}_+ &=& \int_0^{\pi R} ~dy ~e^{k y} g^n_+(y) \Big[ f^m_+(y)-\frac{f^m_+(0)}{g^s(0)}g^s(y) \Big].
\end{eqnarray}
A particularly important result is that for $c>-1/2$, the composite states $\lambda_+^n$ contain no zero mode component, 
\begin{equation}
\omega^{n0}_+=0, \label{nozero}
\end{equation}
since for these $c$ values the source wavefunction (\ref{gsource}) is identical to the zero mode wavefunction (\ref{0mode}). Notice also in this case that $\omega^{s0}_+=1$. We will use these facts often in Sec. 5 when we analyze phenomenology in the holographic basis.

\section{Partial compositeness of SM fermions}
In this section we will determine the explicit elementary/composite content of the standard model fermions in realistic warped models using the holographic basis. As a concrete example we will consider the basic scenario of \cite{custodial}, with the enhanced bulk gauge symmetry which provides custodial isospin to the Higgs sector. In this setup the light fermions are localized on the UV brane, with $c>1/2$, and are primarily elementary, only coupling to the CFT through an irrelevant operator. The heavier fermions, in particular, the left-handed top and bottom quarks and the right-handed top quark, are generally given a bulk mass $c<1/2$, and thus couple to relevant operators. In order to achieve a large top mass, the right-handed top must be significantly IR localized, near $c\sim -1/2$.

Therefore, we will assign characteristic $c$ values and compute precisely the content of the massless eigenstates. To solve the hierarchy problem, we will assume the following values $\pi k R \sim 34.54$ and $k \sim 10^{15} {\rm TeV}$.

\subsection{Light fermions}
For light fermions, such as the electron, the compositeness is completely negligible, as was the case for the graviton \cite{holo}. We can see this explicitly by computing the eigenvector (\ref{ev-}) for $c>1/2$. In this limit the composite coefficient becomes
\begin{equation}
z_n^s\simeq ~b(n,c)~e^{-(c-1/2)\pi k R},
\end{equation}
where $b(n,c)$ is an $O(1)$ coefficient independent of $k$ and $R$.
The holographic basis is essentially identical to the mass eigenbasis, which can be seen in a number of ways. For instance the holographic spectrum (\ref{spectrum2}) yields eigenvalues nearly identical to the true spectrum (\ref{spectrum1}). Also, the transformation matrix (\ref{transform}) is the unit matrix up to exponentially suppressed corrections.

\subsection{Left-handed top and bottom quarks}
We now consider a nontrivial case, that of the left-handed top and bottom quarks $Q_{L3}=\left(t_L , b_L\right)$ where there is appreciable mixing between the sectors. The zero mode is mildly localized on the IR brane, and we will take for concreteness $c=0.4$. Transforming from the holographic basis to the mass eigenbasis, we determine the content of each mode:
\begin{equation}
\left( \begin{array}{c} Q_{L3}^{(0)}  \\ Q_{L3}^{(1)} \\ Q_{L3}^{(2)} \\ \vdots \end{array} \right)
= \left( \begin{array}{rrrr} 
 1       &  -0.484   &  0.290    &  \cdots \\
 0       &   0.874   &  0.200    &  \cdots \\
 0       &  -0.035   &  0.934    &  \cdots \\
\vdots   & \vdots    & \vdots    &  \ddots   
\end{array} \right)
\left( \begin{array}{c} Q_{L3}^{s(1)} \\ Q_{L3}^{CFT(1)} \\ Q_{L3}^{CFT(2)}  \\ \vdots \end{array}\right).
\label{Q3} 
\end{equation}
We see that the zero mode  contains a significant mixture of CFT bound states. Notice also that the massive modes are purely composite. This case has many of the same features as the gauge boson \cite{holo}, which is flat in the bulk and couples marginally to the CFT. 

\subsection{Right-handed top quark}
Consider now the case of the right-handed top quark $t_R$ which is exponentially peaked on the IR brane. Different values are taken in the literature for the mass $c_R$, but in nearly all cases $c_R<1/2$ in order to obtain an $O(1)$ Yukawa coupling. 

If $-1/2<c_R<1/2$, the mixing is  qualitatively similar to that of $Q_{L3}$ just considered. In particular, the zero mode will be mostly elementary, while the KK modes are purely composites. At $c_R\sim -1/2$, the massless mode is approximately half elementary and half CFT bound states. This is
consistent with the scaling dimension of the dual operator (\ref{dimo}), which takes its lowest value at this point. We might have instead guessed that $t_R$ was primarily composite, based on its IR localization. However, as discussed in \cite{schrod}, localization is only a rough guide to the dual interpretation, in particular in the region where strong mixing occurs.

Now consider $c_R<-1/2$. We know that on this branch the source field marries with a new elementary field and becomes massive. On the other hand, there is an ultra-light mode in the CFT spectrum (\ref{lightcft}). We therefore expect the SM $t_R$ to be primarily a composite state. To see this using the holographic basis, let us take a phenomenological value $c=-0.7$, and compute the transformation matrix:
\begin{equation}
\left( \begin{array}{c} t_{R}^{(0)}  \\ t_{R}^{(1)} \\ t_{R}^{(2)} \\ t_{R}^{(3)} \\ \vdots \end{array} \right)
= \left( \begin{array}{rrrrr} 
  0.9796 & \sim -1    &  \sim 0   & \sim 0 &   \cdots \\
  -0.1807 &  \sim 0   & \sim -1     & \sim 0 &   \cdots \\
  0.0511 &  \sim 0   &  \sim 0   & \sim - 1 &   \cdots \\
  -0.0470 &  \sim 0   & \sim 0    & \sim 0 &   \cdots \\
\vdots   & \vdots    & \vdots    & \vdots &   \ddots   
\end{array} \right)
\left( \begin{array}{c} t_{R}^{s} \\ t_{R}^{CFT(1)} \\ t_{R}^{CFT(2)}  \\ t_{R}^{CFT(3)} \\ \vdots \end{array}\right).
\label{tR} 
\end{equation}  
Notice that the SM $t_R$ is roughly a 50/50 mixture of the source field and the first CFT composite state. Also for this case the KK modes now contain some elementary component, different from the case $c>-1/2$.

\section{Phenomenology in the holographic basis}

We will now use the holographic basis to understand the phenomenology of warped models from the dual perspective of SM partial compositeness. Of course, at a qualitative level (and even semi-quantitative), many of these translations have been discussed at various points in the literature. The new aspect that the holographic basis brings is a quantitative interpretation of these 5D geometric successes. In addition, our qualitative understanding is also improved, due to the fact that we can examine the structure of the effective Lagrangian - the quadratic mixing between individual states in the elementary and composite sectors, as well as the various couplings and interactions that these states possess.  We will examine three results: natural Yukawa hierarchies among SM fermions, suppression of FCNC (RS GIM mechanism), and custodial SU(2) and the $T$ parameter.

\subsection{Yukawa coupling hierarchies}

Yukawa coupling hierarchies are elegantly explained in RS models by fermion localization. In the dual picture, fermion localization translates into the dimension of the CFT operator, and thus SM fermion masses result from elementary fields coupling to either relevant or irrelevant CFT operators. We now analyze the Yukawa couplings in the holographic basis emphasizing the connection with operator dimension. 

Let us consider the simplest case of a Higgs boson $H(x)$ confined to the IR brane, corresponding to a purely composite field in the dual picture. The Yukawa interactions between bulk fermions $\Psi_{iL}(x,y), \Psi_{jR}(x,y)$ with bulk masses $c_{iL}$ and $-c_{jR}$, where $i,j$ are flavor indices, are then
\footnote{The negative sign for the bulk mass is simply for convenience, allowing us to use the expressions (\ref{gCFT}) and (\ref{gsource}) for both $\psi_{iL}$ and $\psi_{jR}$.} 
\begin{equation}
S_{Yukawa}=\frac{\lambda_{ij}^5}{k} \int d^5x \sqrt{-g}\Big[\overline{\Psi}_{iL}(x,y)H(x)\Psi_{jR}(x,y) +{\rm h.c.}\Big]\delta(y-\pi R),
\end{equation}
where $\lambda_{ij}^5$ is dimensionless. Canonically normalizing the Higgs action $H\rightarrow e^{\pi k R} H$ and rescaling the fermion fields as usual, we expand the fields in the holographic basis 
(\ref{holo+}) to find the low-energy Yukawa interactions:
\begin{eqnarray}
S_{Yukawa}&=&\lambda_{ij}^5 \int d^4x\Big\{
\left(\zeta_{ss}\overline{\psi}^{s}_{iL} H \psi^{s}_{jR} +{\rm h.c.}\right)\nonumber\\
&+&\sum_n \Big[ \left(\zeta_{sn}\overline{\psi}^{s}_{iL} H \lambda^{n}_{jR} +{\rm h.c.}\right)+\left(\zeta_{ns}\overline{\lambda}^{n}_{iL} H \psi^{s}_{jR} +{\rm h.c.}\right)\Big]\nonumber\\
&+&\sum_{n,m} \left(\zeta_{nm}\overline{\lambda}^{n}_{iL} H \lambda_{jR}^{m} +{\rm h.c.}\right) + \dots
\Big\},
\end{eqnarray}
where the omitted terms do not contain a zero mode component. The overlap integrals $\zeta$ determine the strength of the effective Yukawa interactions:
\begin{eqnarray}
\zeta_{ss}&=&\frac{e^{\pi k R}}{k}g^{s}_{iL}(\pi R)g^{s}_{jR}(\pi  R), \label{yss}\\
\zeta_{ns}&=&\frac{e^{\pi k R}}{k}g^{n}_{iL}(\pi  R)g^{s}_{jR}(\pi  R), \label{yns} \\
\zeta_{sn}&=&\frac{e^{\pi k R}}{k}g^{s}_{iL}(\pi  R)g^{n}_{jR}(\pi  R),  \label{ysn}\\
\zeta_{nm}&=&\frac{e^{\pi k R}}{k}g^{n}_{iL}(\pi  R)g^{m}_{jR}(\pi  R), \label{ynm}
\end{eqnarray}
where the holographic wavefunctions are given in (\ref{gCFT}) and (\ref{gsource}) with the appropriate $c$ values.  

For most phenomenological cases, $c>-1/2$, so that the SM fermions are primarily elementary. More precisely, in this case only the source field $\psi^s(x)$ contains the zero mode $\psi_+^0(x)$, while the composite modes are written in terms of a linear combination of massive KK states (\ref{nozero}). Thus only the strength of the source-source Yukawa coupling (\ref{yss}) is important for SM fermions:
\begin{equation}
\zeta_{ss}=\sqrt{\frac{5-2\Delta_i}{e^{(5-2\Delta_i)\pi k R}-1}}\sqrt{\frac{5-2\Delta_j}{e^{(5-2\Delta_j)\pi k R}-1}}e^{(5-\Delta_i -\Delta_j) \pi k R}.
\end{equation}

\vspace{0.4cm}
\noindent{\bf Light fermions}
\vspace{0.2cm}

Light fermions are described by the case $c>1/2$, and in this case the scaling dimensions of the CFT operators are large $\Delta_{i,j}> 5/2$ (\ref{dimo}). Therefore the source/CFT mixing is irrelevant, and the coupling to the Higgs $\lambda^5_{ij}\zeta_{ss}$ is exponentially suppressed:
\begin{equation}
\zeta_{ss}\simeq\sqrt{(2\Delta_i-5)(2\Delta_j-5)}e^{-(\Delta_i +\Delta_j-5) \pi k R}.
\end{equation}

\vspace{0.4cm}
\noindent{\bf Heavy quarks}
\vspace{0.2cm}

To produce the large masses at low energy for the top and bottom quarks, the elementary fermions must couple to the CFT through a relevant interaction, implying that one or both dual CFT operators have scaling dimension $\Delta_{i,j}<5/2$. In the case of the bottom quark, this can be achieved by having the left-handed bottom $b_L$ slightly IR localized, with ($\Delta_{b_L}<5/2$) and the right-handed bottom $b_R$ UV localized, with ($\Delta_{b_R}>5/2$), yielding the coupling
\begin{equation}
\zeta_{ss}\simeq\sqrt{(5-2\Delta_{b_L})(2\Delta_{b_R}-5)}e^{-(\Delta_{b_R}-5/2) \pi k R}.
\end{equation}
The suppression is not as severe and a larger Yukawa coupling is obtained. The key point is that $b_L$ couples to the CFT through a slightly relevant operator.

For the large top mass both $t_L$ and $t_R$ must interact with the CFT via relevant operators   
($\Delta_{t_L,t_R}<5/2$), and an $O(1)$ Yukawa coupling is obtained:
\begin{equation}
\zeta_{ss} \simeq \sqrt{(5-2\Delta_{t_L})(5-2\Delta_{t_R})}.
\end{equation}

\vspace{0.4cm}
\noindent{\bf Composite fermions}
\vspace{0.2cm}

For heavy quarks, it may also be feasible to have $c<-1/2$, in which case the zero mode becomes mostly composite. In this scenario, both the source field and the composite CFT states contain some zero mode component, and thus in principle all of the couplings (\ref{yss})-(\ref{ynm}) contribute to the SM Yukawa coupling. In particular, in the region $-3/2<c<-1/2$, where the source couples to a relevant operator, this is indeed true and all couplings are important. However, the result is that the Yukawa coupling is still of order one. 

For illustrative purposes, let us examine a more extreme case, which may or may not be relevant for phenomenology, where the heavy fermions are extremely IR localized $c<-3/2$. In this case, the zero mode is almost entirely built from the first composite resonance (see the discussion in Section 3.2). For generic massive composites ($n>1$), the wavefunction (\ref{gCFT}) evaluated at the IR boundary can be approximated to a high degree of accuracy by $g^n_+(\pi R)\simeq \sqrt{2 k}e^{-\pi k R/2}$, and thus the strength is order one, 
\begin{equation}
\zeta_{nm}\simeq 2, \quad\quad\quad n,m>1.
\end{equation}
However, we know that the first composite state becomes exponentially light for $c<-1/2$ (\ref{M0}). The holographic wavefunction for this state $g^1_+(y)$ approximately becomes
\begin{equation}
g^1_+(y)\sim 2\sqrt{(1-2c)k}~e^{(c-1/2)\pi k R} e^{ky/2} \sinh{\left(c+\frac{1}{2}\right)ky},
\end{equation}
which for $y>0$ matches the zero mode wavefunction $f^0(y)$ (\ref{0mode}). Thus, the SM Yukawa coupling in this range of $c$ is enhanced:
\begin{equation}
\zeta_{11} \simeq \sqrt{(1-2c_{iL})(1-2c_{jR})}.
\end{equation}

\subsection{Gauge interactions and FCNC suppression}

\begin{figure}[t]
\centerline{\includegraphics[width=.9\textwidth]{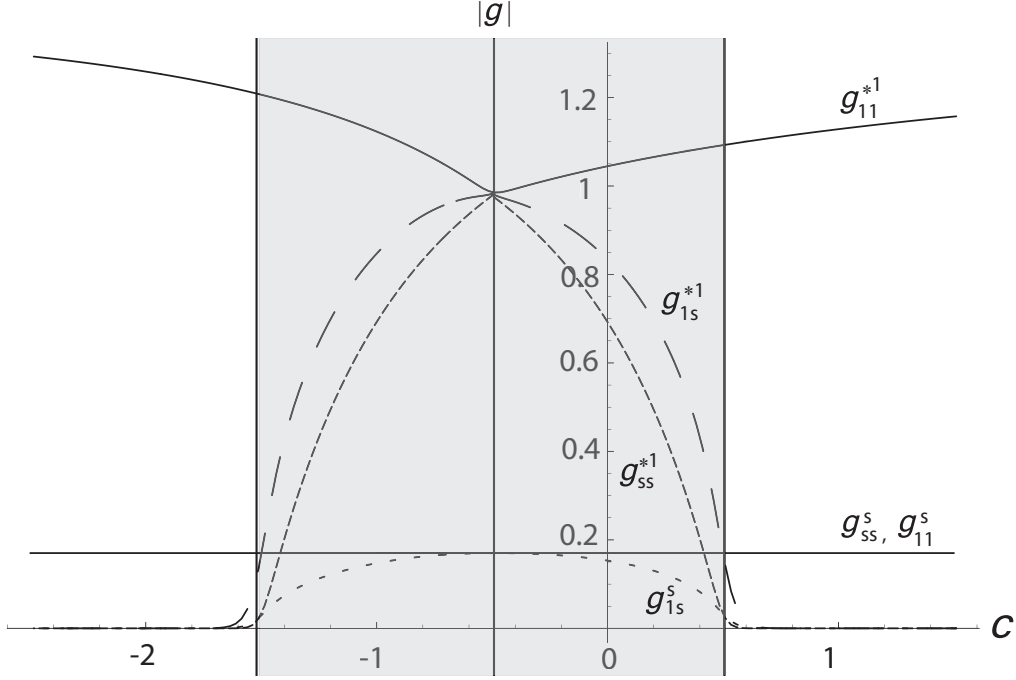}}
\caption{The elementary/composite gauge couplings (magnitudes) for the source field $\psi^s(x)$ and the first composite fermion $\lambda_+^1(x)$ between $-2.5<c<1.5$. The shaded region indicates relevant mixing.}
\label{fig1}
\end{figure}

The holographic basis has given us a quantitative understanding of Yukawa couplings in the dual theory. Similar insight can be gained into the structure of gauge interactions, which we analyze in this section. Consider the following interaction between a bulk gauge field $A_M(x,y)$ and fermion $\Psi(x,y)$:
\begin{equation}
S_{gauge}=g_5\int d^5 x \sqrt{-g} ~\overline\Psi e^M_A \Gamma^A A_M \Psi,
\label{gint}
\end{equation}
where $g_5$ is the 5D gauge coupling. In terms of $\psi_\pm(x,y)$, this becomes
\begin{equation}
S_{gauge}=g_5\int d^5 x \Big[e^{ky}\overline\psi_+ \gamma^\mu A_\mu \psi_+ 
+ e^{ky}\overline\psi_- \gamma^\mu A_\mu \psi_- \Big]~.
\label{gint2}
\end{equation}
Let us focus on the interaction for $\psi_+$, since this is the field we have chosen to contain a massless mode. To decompose this interaction, we must recall the holographic basis for the gauge field \cite{holo}, which is reviewed in the Appendix. 
Using (\ref{App}) and the expansion for $\psi_+(x,y)$ (\ref{holo+}), we have
\begin{eqnarray}
S_{gauge}&=&\int d^4 x \Bigg\{ g^s_{ss} \overline\psi^s\gamma^\mu A^s_\mu \psi^s + \sum_{n=1}^{\infty} g^{\ast n}_{s  s} \overline\psi^s\gamma^\mu A^{ \ast n}_\mu \psi^s\nonumber \\
&& +\sum_{n=1}^{\infty} \Big[
g^s_{nn}\overline\lambda^n_+ \gamma^\mu A^s_\mu \lambda^n_+ 
+ \left(
g^s_{ns} \overline\lambda^n_+ \gamma^\mu A^s_\mu \psi^s + {\rm h.c.}\right) \Big] \nonumber \\
&&
+ \sum_{m,n=1}^{\infty} \left(g^{\ast n}_{m s} \overline\lambda^m_+\gamma^\mu A^{\ast n}_\mu \psi^s+{\rm h.c.}\right) 
+\sum_{l,m,n=1}^{\infty}\left(g^{\ast m}_{l  n} \overline\lambda^l_+\gamma^\mu A^{\ast m}_\mu \lambda^n_+ \right)\Bigg\}~,
\label{gint3}
\end{eqnarray}
where $A_\mu^s(x)$, $A_\mu^{\ast n}(x)$ correspond to the source field and composite CFT states, respectively. Notice there are purely elementary, purely composite, as well as mixed vertices. The 4D couplings are given by wavefunction overlap integrals:
\begin{eqnarray}
g^s_{ss}&=& g~,  \\
g^s_{nn}&=& g~, \\
g^s_{ns}&=& g z_n^s~, \\
g^{\ast n}_{s s}&=&g_5 \int_0^{\pi R} dy ~e^{ky} g^s g^{\ast n} g^s~, \label{gsstars}\\
g^{\ast n}_{m s}&=&g_5 \int_0^{\pi R} dy~ e^{ky}g_+^m g^{\ast n} g^s~,\\
g^{\ast m}_{l n}&=&g_5 \int_0^{\pi R} dy~ e^{ky}g_+^l g^{\ast m} g_+^n~,
\end{eqnarray}
where $g= g_5/\sqrt{\pi R}$ is the 4D coupling, $z_n^s$ is given in (\ref{kmixsource}), and $g^{ \ast n}(y)$ is the composite gauge wavefunction defined in the Appendix. The upper (lower) indices on the 4D couplings refer to the gauge (fermion) field.
We have plotted the magnitude of these couplings for the source field $\psi^s(x)$ and the first composite state $\lambda^1_+(x)$ in Fig.\ref{fig1}. Notice that in general, the composites interact much more strongly than the elementary source, as expected since the composite dynamics are determined from an underlying strongly coupled gauge theory. However, in the range $-3/2 < c < 1/2$, indicated by the shaded region in Fig.\ref{fig1}, the source/CFT mixing is relevant and the elementary sector interacts very strongly with the composite states.

A striking success of warped models with fields in the bulk is the suppression of FCNC among light fermions~\cite{gp}.
Generically, when fermions are separated in the extra dimension, large FCNC can be induced since the SM fermions will couple differently to KK modes of gauge bosons. In warped models, however, light fermions in fact couple universally to KK gauge modes. When the fermions are transformed from flavor eigenstates to mass eigenstates, FCNC are not induced precisely because of this universal coupling, and thus these models enjoy a GIM-like mechanism. 

\begin{figure}
\centerline{\includegraphics[width=.8\textwidth]{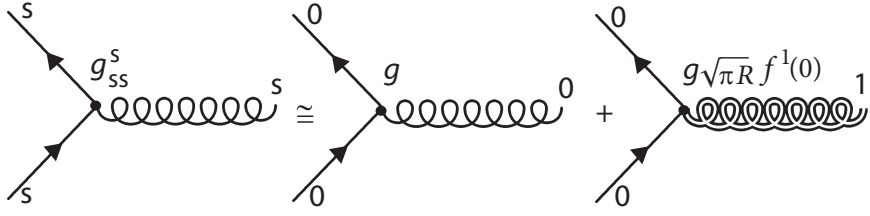}}
\caption{The triple source gauge vertex relevant for light fermions. The double line indicates an excited (KK) mode.}
\label{fig2}
\end{figure}

The approximate GIM mechanism suppressing KK/composite neutral currents has a very straightforward holographic explanation. The light fermions, as we previously discussed, are almost purely elementary, and so for our purposes, the only gauge couplings which are relevant are $g^s_{ss}$ and $g^{\ast n}_{ss}$
\footnote{The couplings $g^s_{nn}$ and $g^{\ast m}_{ln}$, while of order one, involve CFT fields 
$\lambda_+^n$ which do not have any admixture of $\psi_+^0$ for $c>1/2$. See Eq. (\ref{nozero}).}.
Noting the fermion transformation (\ref{sinv}) and the gauge field transformations (\ref{as}) and  (\ref{astar}) in the Appendix, we obtain from (\ref{gint3}) the interaction of SM zero mode fermions $\psi_+^0(x)$ to the first excited gauge mass eigenstate (KK mode) $A_\mu^1(x)$:
\begin{equation}
S_{gauge}\subset \int d^4x \left(g_5 f^1(0)+ \sum_{n=1}^\infty g_{ss}^{\ast n}\omega^{n 1}\right)  \overline\psi^0_+\gamma^\mu A^1_\mu \psi^0_+,
\end{equation}
where $f^1(y)$ is the wavefunction of the first KK gauge mode (mass eigenstate), and the coefficient $\omega^{n 1}$ is given in (\ref{omega}). The coupling between the zero mode fermions and the first gauge KK mode is therefore given by
\begin{equation}
g^1= g_5 f^1(0)+ \sum_{n=1}^\infty g_{ss}^{\ast n}\omega^{n 1}.
\label{g1}
\end{equation}
Notice that the first term in (\ref{g1}) represents a universal contribution to the coupling, $g^1_{universal}=g_5 f^1(0)=g\sqrt{\pi R} f^1(0)$, coming from the triple source vertex. This universal piece is shown in Fig.\ref{fig2}.

The nonuniversal contributions arise from the source fermions coupling to composite vector modes and are contained in the second term in (\ref{g1}). This sum can be performed analytically, as shown in the Appendix, and the result is just what we expect:
\begin{eqnarray}
g^1_{nonuniversal} &=& \sum_{n=1}^\infty g_{ss}^{\ast n}\omega^{n 1} \nonumber \\
& = & g_5 \int_0^{\pi R} dy~ e^{k y} f^0_+(y)f^1(y)f^0_+(y)- g_5 f^1(0). 
\label{g1nutext}
\end{eqnarray}
That is, the nonuniversal contribution is simply the difference between the total coupling and the universal piece, 
$g^1_{nonuniversal}=g^1-g^1_{universal}$.

For light fermions, $c > 1/2$, this universal contribution dominates \cite{flavor2}, and the SM fermions primarily interact through the triple source vertex shown in Fig.\ref{fig2}. This can also be seen in Fig.\ref{fig1}, where $g_{ss}^{\ast 1}$ is suppressed for $c>1/2$. This suppression also occurs for higher couplings $g_{ss}^{\ast n}$. The irrelevant interaction between the CFT and source fermions, a consequence of large anomalous dimensions of CFT operators, means that the direct coupling between SM fermions and composite vector modes is suppressed at low energies. However, the source gauge field couples marginally to a CFT current leading to an ${\cal O}(1)$ universal contribution to $g^1$.

Phenomenologically, it is important to track the flavor-violating contributions to the coupling $g^1$ \cite{flavor2}
\footnote{We thank K. Agashe for discussions on flavor violation, which prompted the expanded analysis in this section.}. In fact, for $c \le 0.6$, the first composite vector provides the dominant contribution to the nonuniversal coupling, $g^1_{nonuniversal} \simeq g_{ss}^{\ast 1}\omega^{11}$, and we can neglect higher modes. It is possible to derive an approximate expression for $g_{ss}^{\ast 1}$ for $c\sim 1/2$:
\begin{eqnarray}
g^{\ast 1}_{s s} \omega^{1 1} & \simeq &  g~\sqrt{2 \pi k R}~\left(\frac{1-2c}{2-2c}\right) e^{-\pi k R}\frac{e^{(2-2c)\pi k R}-1}{e^{(1-2c)\pi k R}-1},  \\
&\simeq& g~\sqrt{2 \pi k R}~\left(\frac{2c-1}{2-2c}\right) e^{(1-2c)\pi k R}, \quad \quad \quad \quad \quad \quad {\rm for} 
\quad c > \frac{1}{2} .
\label{nu}
\end{eqnarray}
Although this expression deviates from $g^1_{nonuniversal}$ for $c>0.6$, fermions with such $c$ values (e.g. the electron) give numerically smaller contributions to flavor-violating operators. Thus, the approximate formula (\ref{nu}) may still be useful in practice. 

For heavy quarks the circumstances are much different for two reasons. First, as we have shown quantitatively in the previous section, such third generation quarks can contain a significant composite mixture, so that in principle every vertex should be accounted for when trying to understand phenomenology. At the same time, these vertices are not suppressed, as shown in Fig.\ref{fig1} for $-3/2<c<1/2$, and can therefore not be neglected. This is also why heavy quarks, like $t_R$, are likely to be important signals of new physics at the LHC, for instance KK gluons~\cite{lhc}, or flavor violation~\cite{flavor}.

\subsection{Custodial protection of the $T$ parameter}
 
Early attempts at extending SM fields into the bulk faced difficulties complying with electroweak precision tests. In particular, excessive contributions to the Peskin-Takeuchi $T$ parameter, arising from the exchange of KK modes of bulk gauge fields were incompatible with a natural solution to the hierarchy problem ($kR\sim 12$)~\cite{TKK}. The key insight into this dilemma was in fact gleaned from holography in Ref.~\cite{custodial}, which realized that the composite Higgs sector lacked the global custodial SU(2) symmetry which normally protects the $T$ parameter. Since a global symmetry of the CFT is dual to a bulk gauge symmetry, Ref. \cite{custodial} remedied the problem by enlarging the bulk gauge symmetry to SU(2)$_L\times$SU(2)$_R\times$U(1)$_{X}$.

The holographic basis allows for an especially transparent understanding of this custodial protection mechanism since we can explicitly analyze the effective Lagrangian. We review the holographic basis for gauge fields in the Appendix. We introduce the 5D gauge fields $W_L^a, W_R^a, X$ corresponding to SU$(2)_L$, SU$(2)_R$, and U$(1)_{X}$ gauge fields, respectively. The symmetry is broken to the SM subgroup on the UV brane via boundary conditions so that only the U$(1)_Y$ hypercharge gauge field $B$ contains a zero mode while the orthogonal combination $Z^{'}$ contains only massive modes. 

Let us decompose the 5D bulk gauge field action in the holographic basis. The fields $W_L^a$ and $B$ contain source fields and thus are expanded according to (\ref{App}), while $W_R^{(1,2)}$ and $Z^{'}$, which contain only composites, are expanded according to (\ref{Amp}). 
This yields the low energy effective Lagrangian:
\begin{equation}
{\cal L} = {\cal L}_{elem}+{\cal L}_{comp}+{\cal L}_{mix}. 
\end{equation}
The elementary sector ${\cal L}_{elem}$ contains the massless source gauge fields $W_L^{as}$ and $B^s$. The composite sector Lagrangian is found to be\footnote{We only keep terms to quadratic order in the Lagrangian, i.e. $W^{a\ast}_{L\mu\nu}$ denotes only the kinetic term, and similarly for the other fields.}    
\begin{eqnarray}
{\cal L}_{comp} & = & -\frac{1}{4}(W^{a\ast}_{L\mu\nu})^2-\frac{1}{2}M^2(W^{a\ast}_{L\mu})^2 \nonumber \\
& & -\frac{1}{4}(W^{a\ast}_{R\mu\nu})^2-\frac{1}{2}M^2(W^{a\ast}_{R\mu})^2 \nonumber \\
& &-\frac{1}{4}(X^{\ast}_{\mu\nu})^2-\frac{1}{2}M^2(X^{\ast}_{\mu})^2,
\end{eqnarray}
where $M$ is the composite mass. The exact SU(2)$_L\times$SU(2)$_R\times$U$(1)_{X}$ global symmetry is manifest in ${\cal L}_{comp}$\footnote{Note that the addition of a bulk mass $\tilde{M}$ for the SU$(2)_R$ gauge boson as in \cite{custodial} corresponds in the dual picture to an explicit breaking of the global symmetry. This would be reflected in ${\cal L}_{comp}$ by the resonances having different masses.}. However the mixing Lagrangian ${\cal L}_{mix}$ contains source/CFT interactions which explicitly break the global symmetry:
\begin{equation}
{\cal L}_{mix}  =  -\frac{\sin{\theta}}{2} W^{as}_{L\mu\nu}W^{\mu\nu a \ast}_L-\frac{\sin{\theta}}{2}B^{s}_{\mu\nu}B^{\mu\nu  \ast}. 
\end{equation}
We expect that the excess contributions to the $T$ parameter are proportional to this breaking. To compute $T$, we first write the system in the mass eigenbasis using the transformation (\ref{tran2by2}). The tree level contribution coming from the dimension 6 operator $t/(4\pi v)^2 |H^{\dag}D_\mu H|^2$ is obtained by integrating out the massive vector resonances \cite{custodial}:
\begin{eqnarray}
t& = &- \frac{128\pi^2} {v^2}(\Pi_{33}(0)-\Pi_{11}(0)) \nonumber\\
& \simeq & 16\pi^2 v^2 g'^2\pi k R\left(\frac{1}{m^2}-\frac{1}{M^2}\right) \nonumber\\
& = & - 16\pi^2  \frac{v^2}{M^2}g'^2 \pi k R \sin^2\theta,
\label{T}
\end{eqnarray}
where $T=t/(8\pi e^2)$, $v$ is the Higgs vacuum expectation value, 
and $m$ is the eigenmass of the $W^a_\mu$ and $B_\mu$ excitations. The second line in (\ref{T}) illustrates the custodial symmetry at work - the second term proportional to $1/M^2$ would be absent without the enlarged SU$(2)_R$ sector, which partially cancels the contribution from the massive SM excitations. This can be explicitly seen by using the relation (\ref{eigmass}), which leaves the remaining nonzero contribution proportional to the strength of the custodial isospin breaking $\sin^2\theta$ as anticipated. Using (\ref{sinth}) the $\pi k R$ enhancement in (\ref{T}) is then cancelled by the source/composite mixing and we qualitatively recover the result of Ref.~\cite{custodial}.   
The holographic basis thus provides a simple understanding of how the cancellation occurs.

\section{Conclusion}

In this paper we have formulated a holographic basis for bulk fermions in a slice of AdS$_5$. Instead of a Kaluza-Klein decomposition, we identify a purely elementary source field and set of CFT bound states in the bulk field expansion. The low-energy theory contains kinetic and mass mixing between the elementary and composite sectors. It is straightforward to transform between the holographic and mass (KK) bases, which tells us quantitatively the elementary and composite field content of the mass eigenstates. As an application we determined the precise admixture of elementary and composite fields which form the SM fermions in warped models.

The holographic basis provides a key addition to the AdS/CFT dictionary applied to warped models. With this work and \cite{holo}, we have shown that such a basis exists for both bosonic and fermionic bulk fields. Any model formulated in a slice of AdS$_5$ has a dual interpretation via the AdS/CFT correspondence, and we hope that our formalism will help quantify, and perhaps simplify, this holographic interpretation. We have illustrated this by examining several phenomenological aspects of warped models, including fermion masses, FCNC, and custodial protection in EWPT. In each of these examples the holographic basis provides improved quantitative and qualitative understanding of the holographic physics of elementary/composite mixing. 

\section*{Acknowledgements}
We thank Kaustubh Agashe for helpful discussions. This work was supported in part by a Department of Energy grant DE-FG02-94ER40823 at the University of Minnesota, and an award from Research Corporation.


\appendix
\def\theequation{\thesection.\arabic{equation}}
\setcounter{equation}{0}

\section{Holographic basis for gauge fields}
Here we present the holographic basis for a bulk gauge field (see \cite{holo} for more details). We will discuss two cases: 1) the general expansion including the entire tower of composite states, and 2) a truncated $2\times 2$ system.

\vspace{0.4cm}
\noindent{\bf General expansion}
\vspace{0.2cm}

First, for a bulk field obeying $(+ +)$ boundary conditions, the holographic basis is given by
\begin{equation}
A_\mu(x,y)
= \frac{1}{\sqrt{\pi R}}A^s_\mu(x) +\sum_{n=1}^{\infty}A^{\ast n}_\mu(x) g^{\ast n}(y),
\label{App}
\end{equation}
The low energy theory consists of a massless source field $A_\mu^s(x)$ and composite resonances $A_\mu^{\ast n}(x)$ with mass $M^{\ast}_n$. The composite wavefunction $g^{\ast n}(y)$ obeys the equation of motion
\begin{equation}
\partial_5 e^{-2ky}\partial_5 g^{\ast n}(y)=-M^{\ast 2}_n g^{\ast n}(y),
\end{equation}
and satisfies $(- +)$ boundary conditions. There is kinetic mixing between $A_\mu^s(x)$ and $A_\mu^{\ast n}(x)$, with strength $z_n^{\ast}$ given by the overlap integral:
\begin{equation}
z_n^{\ast} = \frac{1}{\sqrt{\pi R}}\int_0^{\pi R} dy g^{\ast n}(y). 
\label{zstar}
\end{equation}

Next, we equate the KK (mass eigenstate) expansion with the holographic expansion:
\begin{equation}
\frac{1}{\sqrt{\pi R}}A^0_\mu(x) +\sum_{n=1}^{\infty}A^n_\mu(x) f^n(y) 
= \frac{1}{\sqrt{\pi R}}A^s_\mu(x) +\sum_{n=1}^{\infty}A^{\ast n}_\mu(x) g^{\ast n}(y),
\label{Akkholo}
\end{equation}
where $A^n_\mu(x)$ is the 4D mass eigenstate and $f^n(y)$ is the KK mode wavefunction with (++) boundary conditions. Using the orthogonality of the eigenfunctions, from (\ref{Akkholo}) we can write the source and composite states in terms of mass eigenstates:
\begin{eqnarray}
A_\mu^s(x) & = & A_\mu^0(x)+\sum_{n=1}^\infty \sqrt{\pi R}~f^n(0) A^n_\mu(x), \label{as} \\
A_\mu^{\ast n}(x) &= & \sum_{m=1}^\infty \omega^{nm}A_\mu^{m}(x), \label{astar}
\end{eqnarray}
where the coefficients $\omega^{nm}$ are found to be
\begin{equation}
\omega^{nm}=\int_{0}^{\pi R} dy g^{\ast n}(y) \Big[ f^m(y)-f^m(0) \Big] .
\label{omega}
\end{equation}
Notice only the source field $A_\mu^s(x)$ contains a zero mode gauge field, while the composites $A_\mu^{\ast m}(x)$ are written purely in terms of KK modes since $\omega^{n0}=0$.

For the other relevant case of the bulk field obeying $(- +)$ boundary conditions, the expansion in the holographic basis is simply given by 
\begin{equation}
A_\mu(x,y) = \sum_{n=1}^{\infty}A^{\ast n}_\mu(x) g^{\ast n}(y),
\label{Amp}         
\end{equation}
i.e., there is no source field, but only  composite resonances with masses $M_n^{\ast}$. In this case, the holographic basis is identical to the mass eigenbasis.

\vspace{0.4cm}
\noindent{\bf Truncated system}
\vspace{0.2cm}

Let us specialize to the truncated case of a source field and a single composite mode, $A_\mu^{\ast}(x) \equiv A_\mu^{\ast 1}(x)$. We can derive an approximate expression for the kinetic mixing $\sin \theta \equiv z^{\ast}_1$ (\ref{zstar}):
\begin{equation}
\sin\theta \simeq -\frac{1}{\sqrt{\pi k R}}
\frac{\pi k\,e^{-\pi k R}}{M},
\label{sinth}
\end{equation}
where we have defined $M \equiv M^{\ast}_1$.

The truncated $2\times 2$ system is diagonalized by the nonorthogonal transformation:
\begin{equation}
\left( \begin{array}{c} A_\mu^s  \\  A_\mu^{\ast} \end{array} \right)
=
\left( \begin{array}{cc} 1 & -\tan \theta   \\ 0 & \sec \theta \end{array} \right)
 \left( \begin{array}{c} A^0_\mu   \\ A^1_\mu \end{array} \right). \\
\label{tran2by2}
\end{equation}
After diagonalization, the eigenmass of the excited state (KK mode) is given by
\begin{equation}
m^2=M^2 \sec^2\theta.
\label{eigmass}
\end{equation}

\section{Nonuniversal coupling}
In this Appendix, we compute the nonuniversal contribution to the coupling between SM fermions and the first excited 
gauge KK mode. From (\ref{g1}), we see $g^1_{nonuniversal}$ is given by:
\begin{eqnarray}
g^1_{nonuniversal} &=& \sum_{n=1}^\infty g_{ss}^{\ast n}\omega^{n 1} \nonumber \\
&=& \sum_{n=1}^\infty g_5 \int_0^{\pi R} dy e^{k y} g^s(y)g^{\ast n}(y)g^s(y)  \int_{0}^{\pi R} dy' g^{\ast n}(y') \Big[ f^1(y')-f^1(0) \Big], \nonumber \\
& &
\label{sumnu}
\end{eqnarray}
where we have used (\ref{gsstars}) and (\ref{omega}). We can perform the sum using the completeness of the eigenfunctions:
\begin{equation}
\sum_{n=1}^\infty g^{\ast n}(y)g^{\ast n}(y')=\delta(y-y')~.
\label{complete}
\end{equation}
Using (\ref{complete}) in (\ref{sumnu}), we have 
\begin{equation}
g^1_{nonuniversal} =   g_5 \int_0^{\pi R} dy~ e^{k y} g^s(y)f^1(y)g^s(y)- g_5 f^1(0),
\label{g1nuapp}
\end{equation}
Using the fact that $g^s(y)=f^0_+(y)$ for fermions with bulk masses $c>-1/2$ (see Eqs. (\ref{0mode}) and
(\ref{gsource})), we see that (\ref{g1nuapp}) is equivalent to (\ref{g1nutext}). The first term is the total coupling $g^1$, precisely what one obtains from the usual KK decomposition.


\end{document}